# On the chains of star complexes and superclouds in spiral arms

## Yu.N.Efremov*

*Sternberg Astronomical Institute, MSU, Moscow 119992, Universitetsky pr. 13, Russia*

## ABSTRACT

The relation is studied between occurrence of a regular chain of star complexes and superclouds in a spiral arm, and other properties of the latter. A regular string of star complexes is located in the north-western arm of M31; they have about the same size 0.6 kpc with spacing of 1.1 kpc. Within the same arm segment the regular magnetic field with the wavelength of 2.3 kpc was found by Beck et al. (1989). We noted that this wavelength is twice as large as the spacing between complexes and suggested that they were formed in result of magneto-gravitational instability developed along the arm. In this NW arm, star complexes are located inside the gas-dust lane, whilst in the south-western arm of M31 the gas-dust lane is upstream of the bright and uniform stellar arm. Earlier, evidence for the age gradient has been found in the SW arm. All these are signatures of a spiral shock, which may be associated with unusually large (for M31) pitch-angle of this SW arm segment. Such a shock may prevent the formation of the regular magnetic field, which might explain the absence of star complexes there. Anti-correlation between shock wave signatures and presence of star complexes is observed in spiral arms of a few other galaxies. Regular chains of star complexes/superclouds in spiral arms are rare, which may imply that a rather specific mechanism is involved in their formation, and the most probable one is the Parker-Jeans instability. The spiral pattern of our Galaxy is briefly discussed; it may be of M101 type in the outer parts. The regular bi-modal spacing of HI superclouds is found in Carina and Cygnus (Outer) arms, which may be an indirect evidence for the regular magnetic field along these arms. .

**Key words:**

*E-mail: efremovn@yandex.ru

# 1 INTRODUCTION

The HII regions and OB stars are known to be distributed along the spiral arms of grand design galaxies non-uniformly; rather often they form groupings with sizes of about 0.5–1.0 kpc. Somewhat older objects (such as Cepheids, most of which have ten-fold older age than O-stars) are usually also gathered in the same groups, called star complexes (Efremov 1979). These complexes - the greatest coherent groupings of stars and clusters, which are connected by unity of an origin from the same HI/H$_2$ supercloud (Efremov 1989, 1995; Odekon 2008, de la Fuente Marcos & de la Fuente Marcos 2009, Elmegreen 1994, 2009).

Sometimes the complexes located along an arm at rather regular distances. This is a rare phenomenon; it was discovered by Elmegreen & Elmegreen (1983) who found such chains of complexes in 22 grand design galaxies – amongst of some 200 (B.Elmegreen, private communication) suitable galaxies studied in the Palomar Sky Survey photos. The above authors found the spacing of complexes (which they called HII regions) in studied galaxies to be within 1 – 4 kpc and each string to consist on average of five "HII regions". The gravitational or magneto-gravitational instability developing along the arm was suggested to explain this regularity (Elmegreen & Elmegreen 1983, Elmegreen 1994).

Large groupings of blue stars in the M31 galaxy were first identified by Sidney van den Bergh (1964). He called all these groups OB-associations, suggesting their mean size (about 500 pc) is ten-fold larger than OB associations of the Galaxy because the outer rarefied parts of associations merge with the crowded stellar background of the Milky Way. In fact, by all their characteristic, including their sizes and ages of the oldest stars, about all these groups in M31 should be classified as star complexes; bona fide O-associations (with average sizes of about 80 pc) are concentrated within complexes (Efremov et al., 1987; Efremov 1995).

In irregular galaxies, there is a continuous sequence of star groupings with increasing age (of the oldest stars) and size, starting from clusters to associations to complexes (Efremov & Elmegreen, 1998); in flocculent galaxies the largest complexes transform to short spiral segments (Elmegreen & Efremov, 1996). However, within the spiral arms of grand design galaxies, star complexes are clearly the largest and well determined groupings of the young (and relatively young, such as Cepheids) stars. The connection between star complexes and super associations was discussed elsewhere (Efremov 2004).

Here, we consider the connection between occurrence of a star complex chain in an arm segment and other properties of such an arm. The regularities in spacing of complexes in one of M31 arms which were found in Efremov (2009) are considered in more details, as well as their relation to properties of the magnetic field in the same arm. We also describe the fragmentation of the outer HI arms of our Galaxy into the regularly spaced superclouds, and discuss in short the grand design of the Milky Way spiral pattern. Data on M31 and a few other galaxies let us note that the occurrence of a chain of star complexes along an arm segment may be intrinsically linked with absence of a spiral shock there. We argue that the regular spacing of star complexes and superclouds in a spiral arm, which is a rarely seen pattern, may be causally connected with another rare phenomenon - the regular magnetic field along the arm.

# 2 STAR COMPLEXES IN THE NORTH-WESTERN ARM OF M31

Usual low resolution optical images of M31 display well only one complex (OB78/NGC 206, at the half-way between "SW" and the center of M31 in Fig. 1a). However, in UV-images which show only rather young (up to ~100 Myr) stars – such as taken by the *GALEX* space telescope - the complexes are evident and numerous (Figs. 1c).

They are equally spaced along the inner north-western arm, which is a part of the long arm Baade' (1963) S4 arm. The south-western part of S4 arm (near the major axis) is very bright, it hosts three large segments, which, in fact, can hardly be referred to as "complexes", as we have called them earlier (Efremov et al 1987). Between NW and SW segments of the S4 arm, slightly North of the M31 major axis, there is the above mentioned bright star cloud NGC 206, called association OB78 (van den Berg 1964). In the face-on image this "superassociation" is nothing but a rather long (some 2 kpc) bright segment of the S4 arm; it is completely unlike the much smaller, roundish complexes in the north-western part of the same arm (Fig 1c).

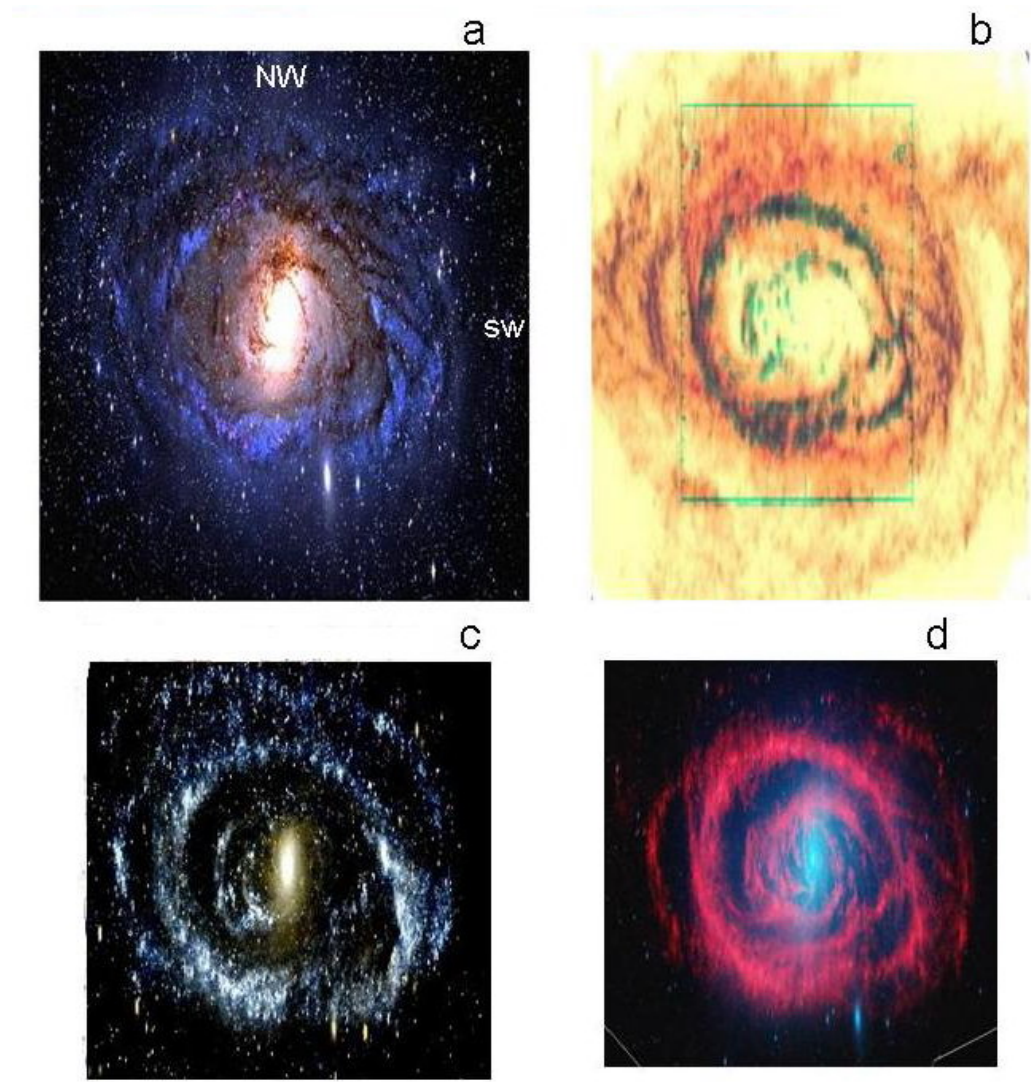

**Figure 1**. Face-on images of the M31 galaxy on the same scale. The angle between the M31 plane and the line of sight is assumed to be 12°, the position angle of the major axis 38° and the distance 690 kpc (Pellet et al. 1978).

    (a) optical image (from the image, obtained by Tony Hallas, with his kind permission)*.
    (b) CO image (from Neininger et al. 2000, green) superimposed on HI image (from Braun et al. 2009, brown).
    *(c) GALEX (Galaxy Evolution Explorer) space telescope UV image**.*
    (d) *Spitzer* space telescope IR image***.

-----------------------------------------------------------------
 *) http://antwrp.gsfc.nasa.gov/apod/ap080124.html
 **) http://www.galex.caltech.edu/media/glx2008-01f_img01.html
**\*\*\***) http://gallery.spitzer.caltech.edu/Imagegallery/image.php?image_name=ssc2006-14a.

The close local correspondence between the lanes of HI, CO and warm dust/PAH molecules (seen in far-IR) is evident in Figs. 1b and 1d, as well as their good all-galaxy agreement, but no certain spiral structure is seen. Instead, there is a complicated pattern with spiral arms, arcs and spurs, resulting evidently from M31 plan corrugations and interactions with M33 and dwarf neighbors. Note also holes in gas/dust around O-star concentrations (the largest one is around OB78/NGC 206). This pattern was discussed many times, and our present topic is the structure of spiral arms and not the spiral structure of M31. We will try to explain a drastic difference between certain segments of the same arm S4.

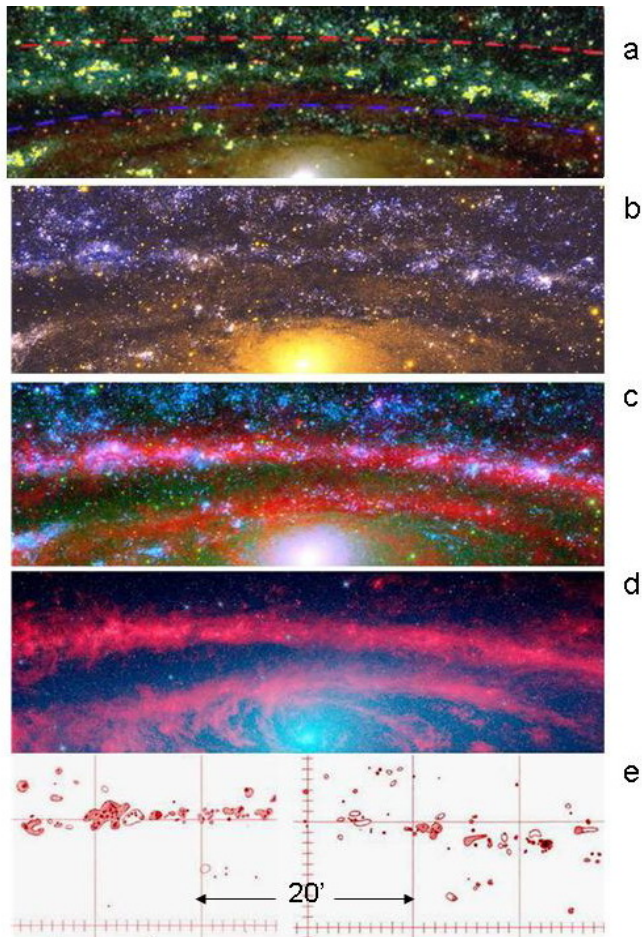

**Figure 2.** Images of the central NW segment of M 31, reduced to the same scale.
10' = 2 kpc along the major axis.
(a) star formation regions (yellow), detected on the basis of the *GALEX* data (detail of fig. 13 in Kang, Bianchi & Rey, 2009).
(b) *GALEX* far-UV image *).
(c) the *GALEX* image superimposed on the *Spitzer* one. "This image is a false color composite comprised of data from Galaxy Evolution Explorer's far-ultraviolet detector (blue), near-ultraviolet detector (green), and Spitzer's multiband imaging photometer at 24 microns (red)"**).
(d) *Spitzer* telescope far-IR image. Note holes in the warm dust, corresponding to the O-star concentrations and HII regions (compare to Fig. 2c and 2e). Red is 8 micron emission, and blue are stars at 3.6 and 4.5 microns **).
(e) HII regions map (from Pellet et al. 1978).
-----------------------------------------------------------------
*) http://www.galex.caltech.edu/media/glx2008-01f_img01.htm)
**) http://gallery.spitzer.caltech.edu/Imagegallery/image.php?image_name=ssc2006-14a

Let us consider the NW part of the S4 arm in more details. It is shown in Fig. 2 and 3 at different wavelengths. Isolated star complexes are clearly visible in *GALEX* telescopes far-UV images and more so - in the overlay of the *GALEX* and *Spitzer* (far-IR) images, taken from the *Spitzer* telescope site. The regular spacing of the complexes along this segment of Baade's S4 arm is now quite evident, as was noted in Efremov (2009). The concentration of HII regions inside the star complexes is evident too, at least in the right-hand part of Fig. 2e.

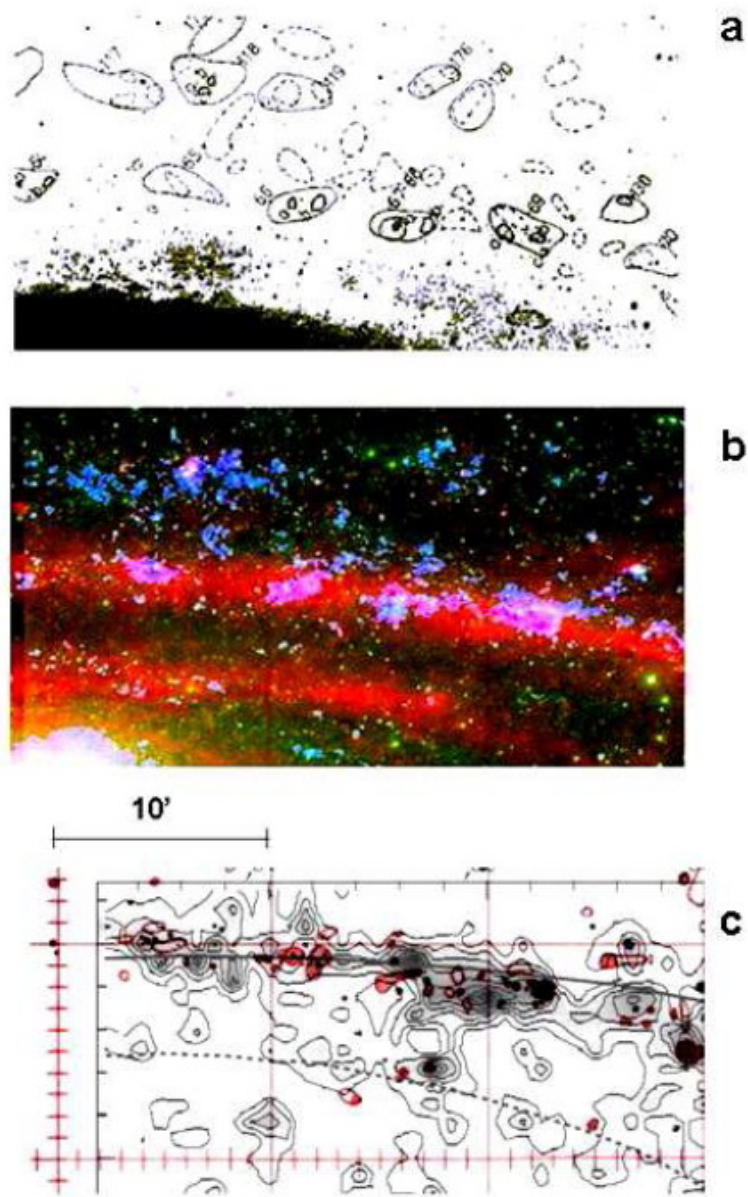

**Figure 3.** Star complexes, CO gas and HII regions in the segment of the north-western arm, located to SW of the M31 minor axis.

(a) Star complexes outlined by blue stars, resolved in the UBV plates obtained with 2-m Rozhen observatory reflector (Efremov et al. 1987).

(b) Enhanced version of the *GALEX* and *Spitzer* telescope overlap image.

(c) HII regions (Pellet et al. 1978, the most bright regions are black ones) and CO gas isolines (Loinard et al. 1999). Contrary to HII regions, CO clouds correspond to star complexes only in general, which is evidently due to holes in gaseous medium, produced by O-stars/HII regions.

As the Fig. 3 demonstrates, complexes seen in *GALEX* images are closely matched to those outlined by Efremov et al. (1987) by eye. The above authors the large-scale UBV plates taken with 2-m reflector of Rozhen Observatory (Bulgaria), with M31 resolved into stars (Fig. 3a). The resolution of these plates is much higher than of *GALEX* images. Most of complexes found with the Rozhen telescope plates agree well with van den Bergh's associations or their groupings, and we have preserved van den Bergh's (1964) designations.

## 3  STAR COMPLEXES AND MAGNETIC FIELD

Along the segment of the NW arm, shown in Fig. 4, Beck et al. (1989) have found the strong polarization at the wavelengths of 20.1 and 6.3 cm, and the regular wavy magnetic field along the arm. We noted (as Figs. 2 and 3 demonstrated) the really striking regularity of spacing of complexes in the part of the NW arm, located to SW (to right-hand side) of the M31 minor axis. Moreover, the sizes of complexes are quite similar there. Now, it looks like the correlation is seen in Fig. 4 between positions of complexes and properties of the magnetic field.

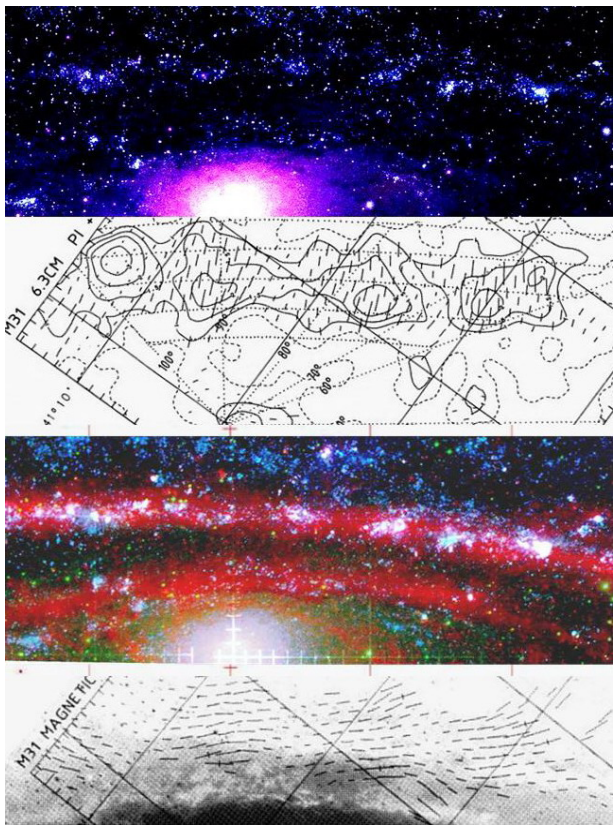

**Figure 4**. The star complexes, warm dust and the magnetic field vectors along the NW arm. The scale and positions along X –coordinate are the same.
From upper to bottom panels:
    1) *Swift* space telescope far-UV image*.
    2) Linearly polarized emission at λ6.3 cm. The lengths of the E-vectors show the degree of polarization. Azimuthal angles θ in the M31 plane are shown (detail of fig. 4 in Beck et al. 1989).
    3) GALEX\Spitzer image superimposed on the map of the HII regions (white, from Pellet et al. 1978).
    4) Orientation of the magnetic field ($\chi_B$). The lengths of the vectors indicate the degree of linear polarization at λ6.3 cm (detail of fig. 6 in Beck et al. 1989).

In this arm segment (within the range of galactocentric distances 8–10 kpc) Beck et al. (1989, p. 63) found the wavy variation in the degree of polarization at λ6.3 cm, the wavelength being 2.3 kpc (Figs. 5). Assuming again the same plane inclination 12° (as Beck et al. 1989 have done) we obtain from Fig. 5 the average spacing 1.13 +\- 0.15 kpc for five complexes (OB64, 65, 66\67, 68, 69 – see Fig. 3) in the 58 -94° interval of azimuthal angles. Their average size is 0.64 +\- 0.16 kpc (all values are measured along the arm). Thus, the spacing between these five complexes along this arm segment corresponds to half the wavelength of the magnetic field. These complexes appear to have a tendency to be located mostly near either the maxima or minima of polarization (Fig. 5).

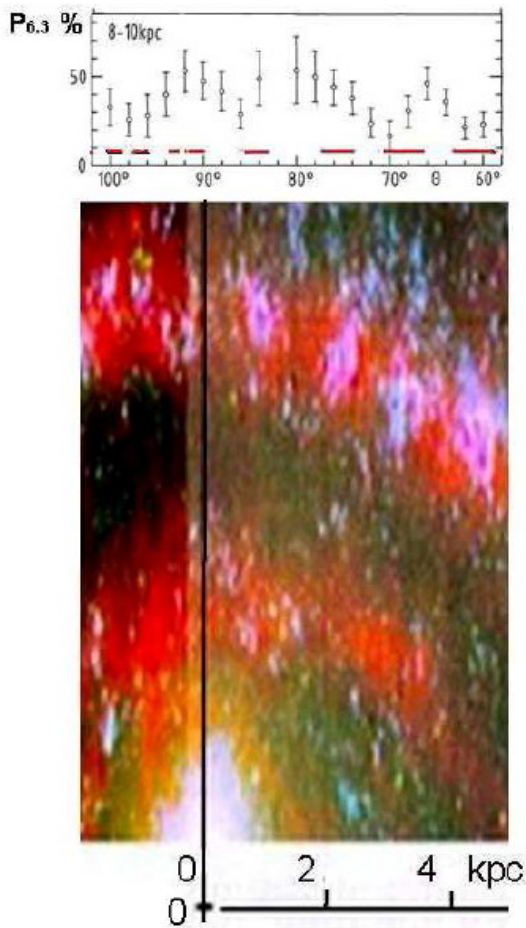

**Figure 5.** Star complexes and magnetic field in the segment of the north-western arm. Upper panel: Degree of polarization at λ6.3 cm (%) averaged in 2° wide azimuth angle sectors in the plane of the galaxy (according to fig. 8 of Beck et al. 1989), and azimuth angles of star complexes (shown as the red lengths). The azimuth angles in the M31 plane are 0° along the SW major axis and 90° along the minor NW axis.
Bottom panel: the face-on *GALEX/Spitzer* image of the respective part of M31. The center of M31 is at (0, 0).

Beck et al. (1989) concluded that the variations of the polarization parameters along the NW arm are indicative of a three-dimensional wavy magnetic field structure, probably------ related to the Parker–Jeans (PJ) instability. We may now suggest that formation of star complexes at distances that are half the wavelength of the field variation along the arm, hosting these complexes, is also related to this instability.

---------------------------------------------------------
*) http://swift.gsfc.nasa.gov/docs/swift/swiftsc.html

Elmegreen (1982) found that complexes formed under action of the PJ instability should have the preferred mass (and therefore, the size too). A number of recent theoretical investigations ( Franco et. al. 2002, Lee et al. 2004, Lee & Hong 2007, Mouschovias et al. 2009) demonstrated arising of the regular spacing of star-forming complexes, resulted due to this instability. Now we indeed see this in the Andromeda galaxy - mostly due to the *GALEX* telescope, which show the stars of appropriate age only, concentrating in star complexes. Before these data have appeared, nobody has noted this striking regularity.

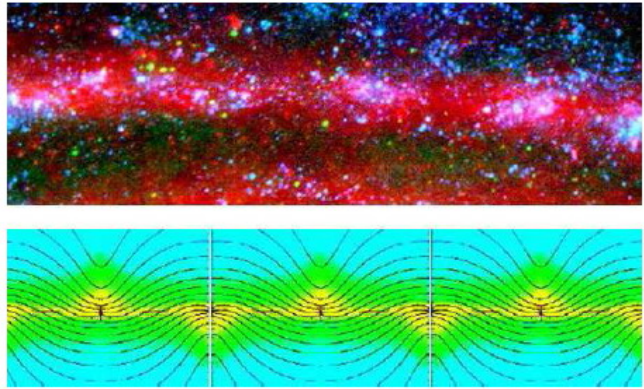

**Figure 6.** Morphological similarity between the structure of the M31 NW arm and a theoretical pattern of the PJ instability along an arm.
Upper panel: the observed warm dust (red) and stellar (blue) structure of the M31 NW arm.
Bottom panel: the theoretical pattern of gas (green), denser star-forming gas (yellow), and magnetic field lines (enhanced and repeated detail of fig. 3 in Mouschovias et al 2009).

It is possible that the wavy structure of the gas/dust lane, seen in Figs. 2b, 2c, 4 and 6, may be the direct optical evidence of the gas lane shaped under action of the PJ instability. The wavy appearance of this lane and complexes location there, may probably be explained by their (small) deviations from the M31 average plan, which aroused due to magneto-gravitational instability along an arm (Fig. 6). If so, this is the only case when the large inclination of the M31 plane is useful…

Situation in the north-western arm of M31 seems to be described by the model of the PJ instability, obtained by Franco et al (2002). These authors found that superclouds (having to produce star complexes) located alternatively above and below the mid-plane. The superclouds emerge as the undular mode of the Parker–Jeans instability, developing along the arm, with the spacing 1–1.3 kpc.

In the face-on image of M31, obtained with its plane tilt of $12°$ (adopted by Beck et al. 1989) the complexes appear elongated in the cross-arm direction (Fig. 5), but they transform to about round ones after assuming inclination $16°$ (compare Fig. 7a with 7c). The latter value has recently been inferred from HI data for the positions in the M31 plane, corresponding to this arm segment position (Chemin et al. 2009). The M31 disk plane is well known to be non-planar. Loinard et al. (1999) found from the kinematics and positions of CO clouds that the tilt of the M31 disk varies with Y, from 15° at Y = -10' to 11° at Y = +10'. Here Y = -10' corresponds to the position of the NW arm and the local tilt 15° disproves the possibility that the complexes in the NW arm might be seen inside the gas/dust lane only because the local tilt of the M31 plant might be closer to 0° there.

Alternatively, the elongated shapes of complexes in the Figs.5 and 7a might be due to the appropriate (and ordered) tilt of each complex plane to the local M31 plane. This explanation seems to be farfetched, although local inclinations of some complexes are known in our Galaxy.

To conclude this Section, we note that the rather regular spacing of star complexes can also be seen in the left-hand part of the NW arm (Fig. 2). Unfortunately, no magnetic-field data, similar to those discussed above, available for this portion of the arm. Such data would be of crucial importance.

## 4 STRUCTURE OF THE M31 SOUTH-WESTERN ARM

Walter Baade (1963), who was the first to establish the concentration of HII regions along the spiral arms of the Andromeda galaxy, noted that a spiral arm behaved similarly to a chameleon: in some segments it was filled with stars, in others it might be identified only on presence of the dust clouds. He meant mostly the arm which he designated as S4, the parts of which were our north-western and south-western arms (note the different designation of the first arm used in Beck et al.,1989).

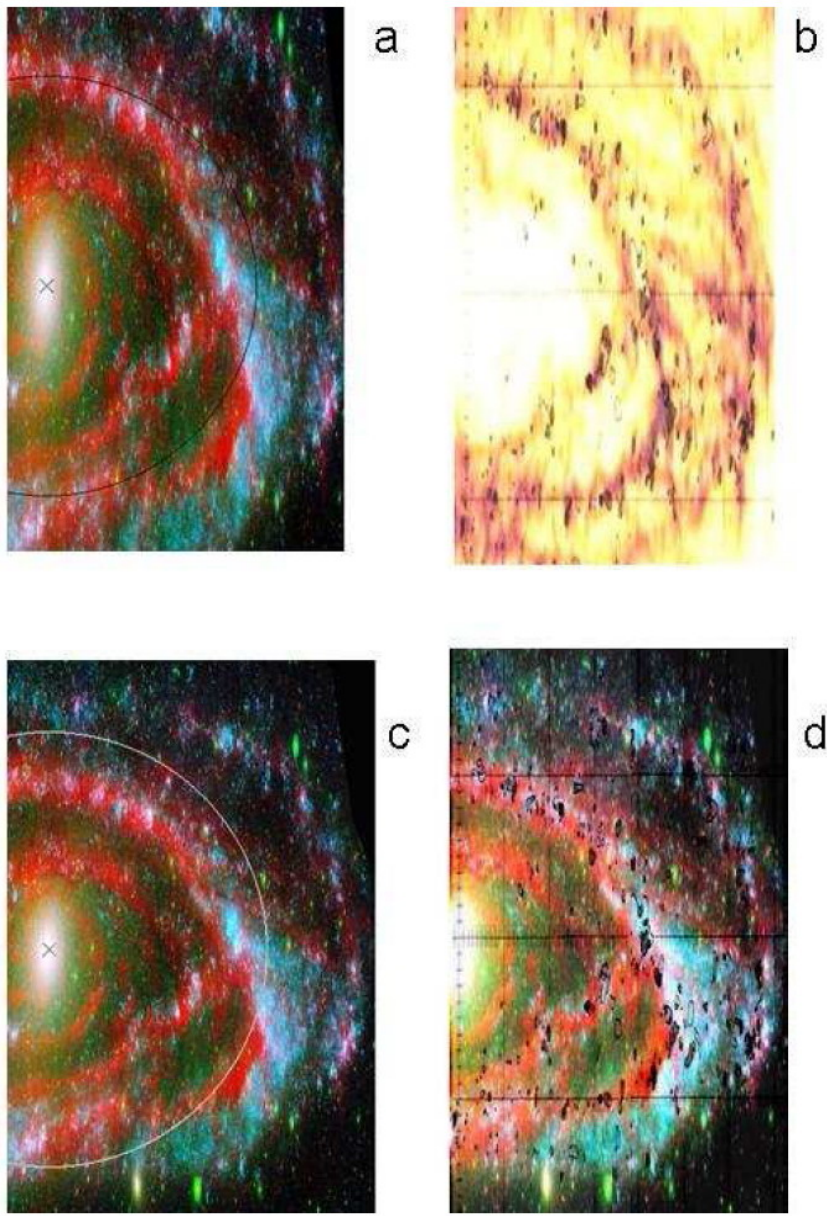

**Figure 7.** The face-on images of M31. The north-west is to the top side, the south-west is to the right-hand side.

(a) *GALEX + Spitzer* telescope transformed image, with the circle added around the galaxy center through NW and SW arm segments. The plan inclination 12°.
(b) HI image transformed from Braun et al. (2009), overlaid on transformed HII map from Pellet et al. (1978). The plane inclination 12°
(c) The same as (a), but the plane inclination 16°
(d) *GALEX + Spitzer* telescope transformed image, overlaid on transformed HII map from Pellet et al. (1978), the plane inclination 16°

In Fig. 7 a drastic difference is seen in the structures of the north-western and south-western parts of the S4 arm. In the NW segment of this arm, the well-separated complexes are located inside the gas–dust arm; in the SW segment, individual complexes cannot be identified at all (discussion on the star cloud OB78 was in Section 2), while the gas/dust lane lies upstream the bright stellar arm. In this south-western arm (often called simply arm S4), where H II regions mostly located between the neutral gas (dust) lane and the stellar arm, the gradient in stellar ages is observed across the arm (Efremov 1985, 1989). The properties of this arm segment are in complete agreement with the density wave theory of spiral arms, which is not a common case.

A possible explanation for this correspondence to the theoretical expectations is as following. We see from Figs. 7a and 7c (where a circle around the galaxy center is shown), that the bright blue segment of the S4 arm (densely populated by early-type high-luminosity stars) has the pitch-angle $i$, unusually large for M31 arms (about 25-30°). The same pitch-angle was found (Efremov 1985) from the angle (in the sky plane image) between the M31 major axis and the arm turning point. The relation between these angles, suggested by Stock (1955), was used. We also see that for the north-western arm and, in general, for almost all of other arms of M31, the pitch-angle is close to zero. According to the classical theory (Roberts et al. 1975), the degree of gas compression by a spiral shock wave is determined by the component ($W$) of difference between the velocities of solid-body rotation of the density wave ($V$dw), and of differential rotation of the galaxy's gas ($V$) around its center, which (component $W$) is directed perpendicularly to the wave front, i.e., to the inner boundary of the arm:

$$W = (V - V\text{dw}) \sin i$$

This implies that at the same distance from the corotation (where $V = Vdw$), the shock wave should be stronger in the arm segment with the larger pitch-angle. The shock wave compresses gas clouds; the higher the density of the initial gas cloud, the higher is the effectivity of star formation). The large pitch-angle of the arm segment in question leads to a high value of $W$. The gas density increases after passage of the shock wave as the ratio of $W^2$ to $V_T$, where $V_T$ is the isothermal sound velocity, which is close to the dispersion of the gas velocities (Kaplan & Pikelner 1979). Such a strong compression of gas clouds by spiral shock wave is surely able to trigger star formation (e.g. Tamburro et al. 2008). The luminous stars borne in the wave are getting away from their birth places, moving across the arm faster than the spiral pattern (inside the corotation radius). This leads to the age gradient across the segment of the arm in question, which gives the local difference between velocities of the spiral pattern and the differential galaxy rotation. Such a gradient is really observed (Boeshaar & Hodge 1977, Egusa et al. 2009, Martínez-García et al. 2009, Grosbol & Dottori 2009), but in rather rare occasions, probably because the really high value of $W$ (and therefore the gas pressure) is needed to enhance star formation rate. Indeed, the age gradient was found in this south-western segment of S4 arm, from the decline of Cepheid periods (Efremov, 1983, 1989) and luminosities of the brightest stars (Efremov & Ivanov 1982, Efremov 1985) with increasing of distance from the inner edge of the stellar arm (Fig. 8).

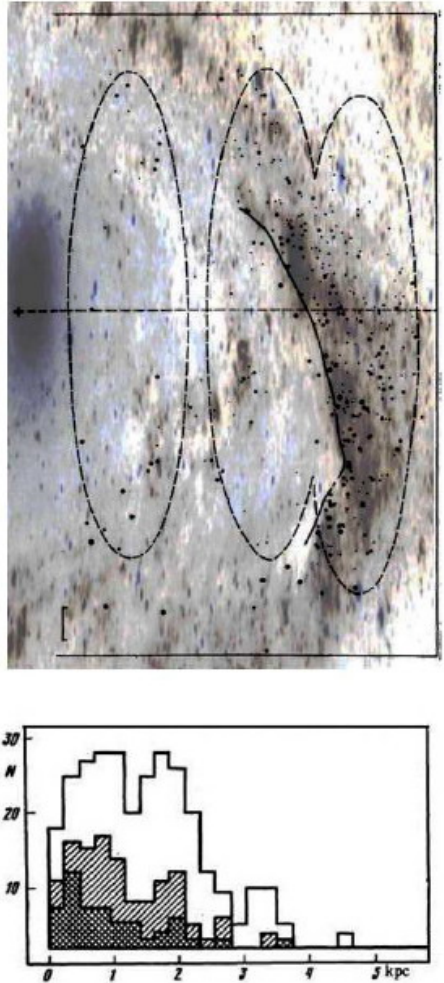

**Figure 8**. Cepheids in the Baade fields of the Andromeda galaxy (the face-on map with the inclination 12°).

Upper panel: Baade's field I, II and III, where Cepheids (black points) were investigated (for details see Efremov 1989, p. 135) overlaid on the *GALEX* image. The inner edge of the S4 arm is shown. The vertical length in the low left corner is equal to 1 kpc.

Bottom panel: distribution of Cepheid distances from the S4 arm edge (from fig. 41b in Efremov 1989). Upper histogram – all the periods, median – periods longer than 10 days, bottom histogram - periods longer than 15 days. According the period-age relation, the longer period the younger a Cepheid (Efremov 2003). Note that these ages do not suffer from the light extinction.

The presence of a shock in this segment of the spiral arm is also confirmed by its rectilinear shape (Efremov 2001), which is immediately apparent in Figs. 7 and 8. According to Chernin (1999) and Chernin et al. (2001), the presence of rectilinear segments in the spiral arms of galaxies is explained by the tendency of a shock wave to straighten out its front. Berdnikov & Chernin (1999) found such segments in the Car – Sgr arm of our Galaxy, too.

# 5 CHAINS OF SUPERCLOUDS IN SPIRAL ARMS OF OUR GALAXY

The regular spacing of HI superclouds and their cores, the giant molecular clouds (GMC), along the Carina arm, was evident in Grabelsky's et al. (1987, 1988) figures, though was not stressed by these authors. It is well seen in Fig. 9a. The HI superclouds were first detected in the outer arms of the Milky Way galaxy by McGee and Milton (1964). The mean mass of these superclouds was found to be $10^7$ solar masses.

Efremov (1998) has attempted to construct a sketch of the MW spiral arms from the data on the supercloud locations. The distances for the superclouds in the Carina arm (IV quadrant) were taken as distances of their GMC cores, given by Grabelsky et al. (1988), the data for superclouds in the I quadrant were taken from Elmegreen & Elmegreen (1987). The 40 kpc long Car – Sgr arm with the pitch-angle 10-12º was the result, with the position similar to one obtained by Grabelsky et al. (1988) from the GMCs. This arm is evidently a notable part of the Milky Way grand design and one may assume that the other arms should be at symmetrical positions. After rotating this arm (assuming the pitch-angle 12º) by 90º around the center, one obtains the Fig. 9b (Efremov 1998, 2009) which exactly corresponds to the Galaxy spiral arms positions (Fig. 9c), outlined by Vallée (2008a) from other data, as it is seen in Fig. 9c.

This four-armed scheme is also in a good agreement with the most recent plan of the MW disk structures (Fig. 9d), suggested by Churchwell et al. (2009). These authors found that Carina and Outer (Cygnus) arms are unseen in the old star distribution. At any rate, just these two arms are the most strong ones in the HI distribution, as Fig. 10a demonstrates.

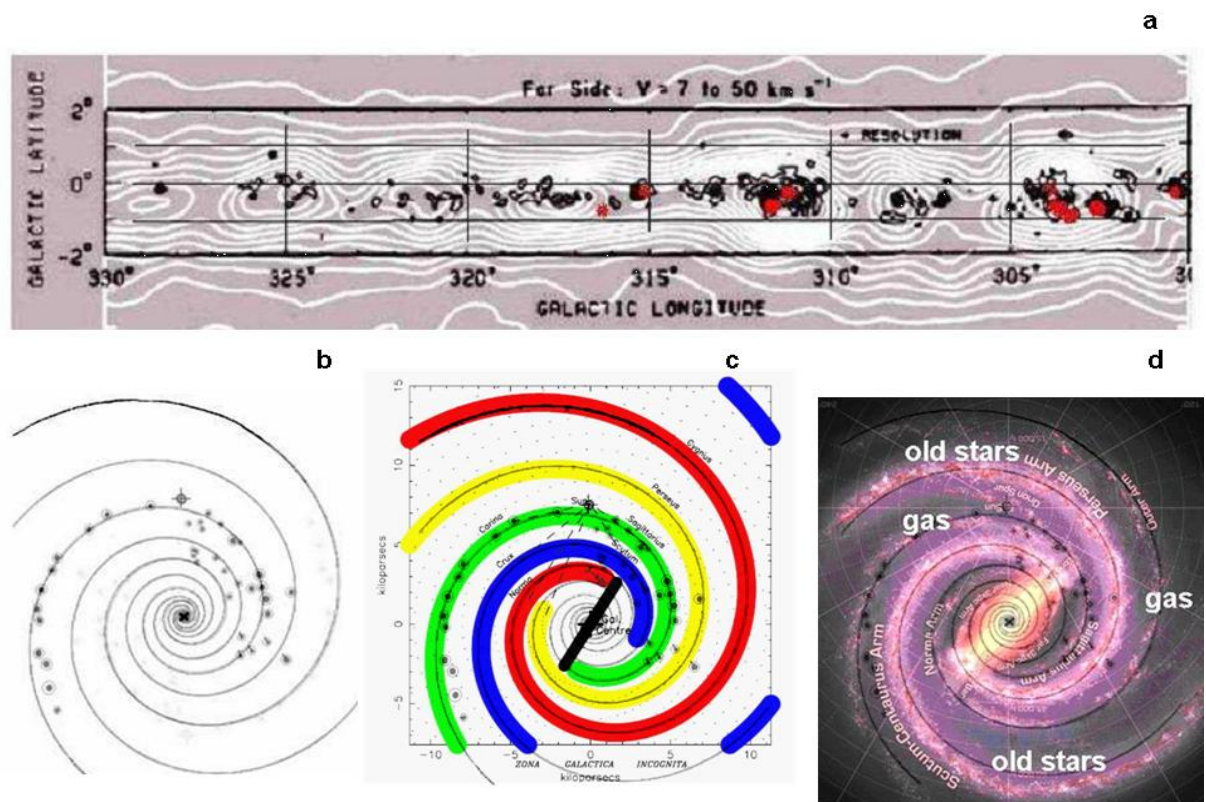

**Figure 9.** HI superclouds in the Carina arm and the Grand Design of the Milky Way.
(a) The composite map of HI (white isolines), CO (black) clouds and HII (red spots) regions in the part of the Carina arm. Data are taken from Grabelsky et al. (1987) and Hou et al. (2009). The radial velocities for all objects are 7 to 50 km/s.
(b) HI superclouds positions in the Carina arm and in the I quadrant. The spiral arm with the pitch-angle 12° is drawn through these superclouds and then is repeated three-fold with turning by 90°.
(c) Fig. 9b superimposed on Vallée's (2008a) sketch of the Milky Way spiral arms.
(d) Fig. 9b superimposed on Churchwell's et al. (2009) sketch of the Milky Way disk structure. .

We noted that both HI strong arms are fragmented into HI superclouds (Fig. 10, upper panel). Their spacing along the arms have mostly values either 0.1 or 0.2 in units of the Sun distance to the center (Fig. 10, middle panel), and we guess that such a bi-modality may be due to formation of superclouds/star complexes at distances, equal sometimes to the full wave-length of the wavy magnetic field, sometimes to the half-wavelength (the latter is the case for the M31 NW arm). A similar situation – absence of a complex between two ones, spacing of which being twice larger than usual - was described for some galaxies by Elmegreen & Elmegreen (1983). They noted that at the position of a missed complex there was still a smaller HII region. It is the case also for the Milky Way: the lonely small CO cloud together with a HII region is at the position of the missed HI supercloud at L = 315° ( Fig. 9a). Note also that each of two deviating points in Fig. 10 (middle) correspond to spacing twice as large as one of the preferred spacing values.

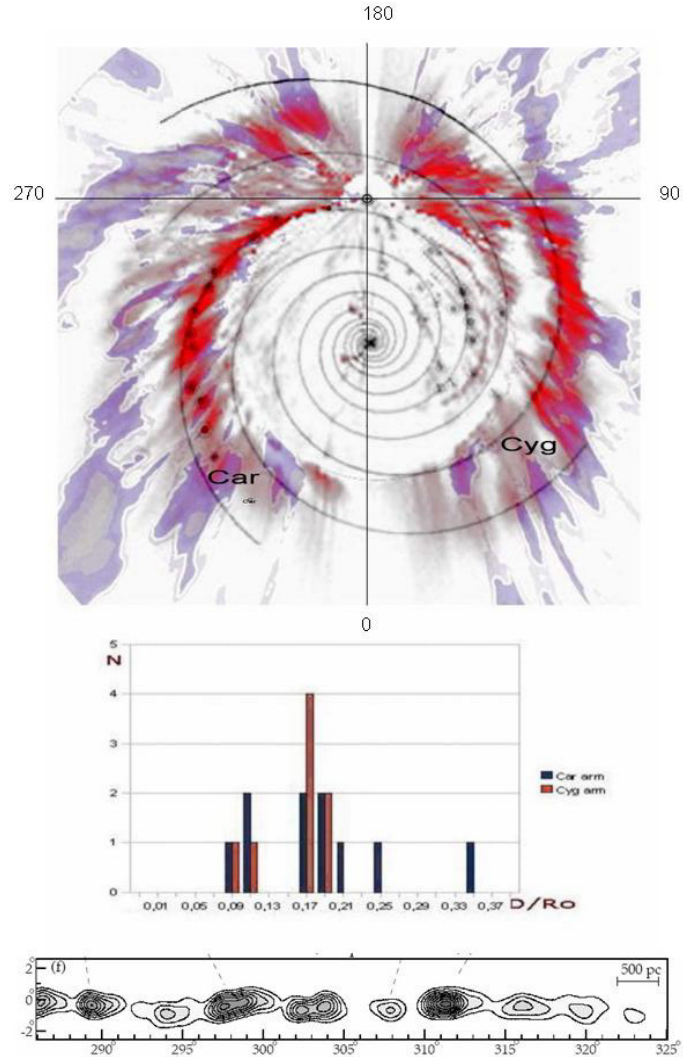

**Figure 10.** Fragmentation of HI arms of the outer Galaxy into superclouds.

Upper panel: edited Nakanishi & Sofue's (2003) figure for distribution of HI in the Galactic plane superimposed on Levine's et al. (2006) one, and on the Fig. 9b. The most dense HI areas are red. These areas, fragmented into superclouds, outline the Cygnus (Outer) arm in the I quadrant, and the Carina arm in the IV quadrant.

Middle panel: Distribution of HI supercloud spacing in Carina and Cygnus (Outer) arms, obtained from the upper panel. Spacing is given in units of the Sun distance to the center.

Bottom panel: The GMCs along the segment of the Carina arm. It is the fig. 6f from Loinard et al. (1999), where the linear scale along all the arm length is about the same and adjusted to the Carina arm average distance of 11 kpc.

This might be considered a support for the possibility of the regular magnetic field along the Milky Way arms. Also, one may note that the chains of clouds, seen in Fig. 9a, and in the bottom panel of Fig. 10 are somewhat similar to the wavy pattern of the M31 northwestern arm. It may be strange enough to see the best regularities in arms of our Galaxy and in the nearest grand design galaxy…

Both complexes of young clusters studied by Alfaro et al. (1992) in the Carina – Sagittarius arm deviate by ~30–40 pc from the Galactic plane to the same direction. If these deviations relate to their origin due to magneto-gravitational instability, their spacing (2.4 kpc) could then correspond to the wave-length of the magnetic field along the arm. Data on the magnetic field in the Galaxy spiral arms are still inconclusive, but the regular field seems to be possible (Vallée 2008b).

The outer parts of the HI arms in Fig. 10a seem to be composed of the rectilinear segments. (The less dense ones, shown in blue, are seen in Levine et al. (2006) figure only, because these authors have imaged the densities above the *local* average values; otherwise the agreement with Nakanishi & Sofue (2003) image is perfect). Such long and straight asymmetric arms are well known in the outer parts of the M101 and NGC 1232 galaxies. Far from a galaxy center such straight arm segments hardly might be due to shock waves, assumed by Chernin (1999). The knee arms might be formed by the gravitational instability of galactic disks,and if so, they are transient features (Clarke & Gittins 2006; Dobbs & Bonnell 2008). The transverse profile of such arms should be symmetric about the gravitational potential well ("gorge"), which means that splitting along an arm into gas and stars should be absent, unlike the density wave spiral arms (far from corotation).

If the outer arms of our Galaxy belong to this type, it is useless to search for the age gradient across them. It follows also that kinematical distances based on observed velocities and the rotation curve may have large errors. This may well be the case, considering the recent data on the very large deviations of the velocities of star formation regions (SFRs) from expected in the density wave theory. Baba et al. (2009) have concluded from the velocity data that this theory does not explains the Galaxy arms, which should be instead of a transient type. However, the inner arms of the Galaxy may well be the density waves, like it is the case for M101 and NGC 1232 galaxy. Note that most of large velocities for SFRs were found far from the MW center.

# 6 STAR COMPLEXES AND DUST LANES

Adopting the dark lane along the stellar arm as a signature of the spiral shock wave (though the strict coincidence of a shock and a lane is not obligatory - Gittins & Clarke 2004), we may expect that a regular chain of complexes may be connected with the absence of an upstream dust lane, like it the case in M31.

The mutual avoidance of the complexes (or at least chains of complexes) and the shock waves seems to be seen in the images of several galaxies. For example, in M74 no individual star complexes can be identified in the shorter (inner) arm, with the dust lane running in front of its inner edge, while there are seven widely separated complexes in the other, longer arm. They are especially bright in the far-UV image, obtained with UIT (Fig. 11). The regular spacing of complexes within the latter was found by Elmegreen & Elmegreen (1983). Now we stress that there is no dust lane running upstream this stellar arm inner edge, but in the middle of it there is a (warm) dust lane, connecting these complexes. This is well seen in the infrared image obtained by the *Spitzer* telescope, recording warm dust (Fig. 11c): the situation is rather similar to one observed in the M31 NW arm. Unfortunately, there is no published magnetic field data for M74.

B.Elmegreen et al. (2006) studied the size distribution for young stellar groupings in M74 within the galaxy central region. They found the cumulative size distribution to be a power law and the luminosity distribution is approximately a power law as well. Their conclusion was that these results suggest a scale-free nature for stellar aggregates in this galaxy. However the studied region included only the galaxy center with the shorter arm, but not encompassed the outer part of the longer arm with its regularly spaced star complexes. This longer arm (Fig.11) should give excess of the large size aggregates at least outside the corotation radius, where no spiral shock is expected. The value of this radius, found by Egusa et al. (2009) implies the chain of complexes is outside it, indeed. Recently Sanchez Gil et al. (E.Alfaro, private communication) detected the age gradient just across the shorter arm, whilst no gradient was found in the longer one. These findings demonstrate once more that the shock wave signatures and complexes at regular spacing are mutually excluding features in a given arm segment.

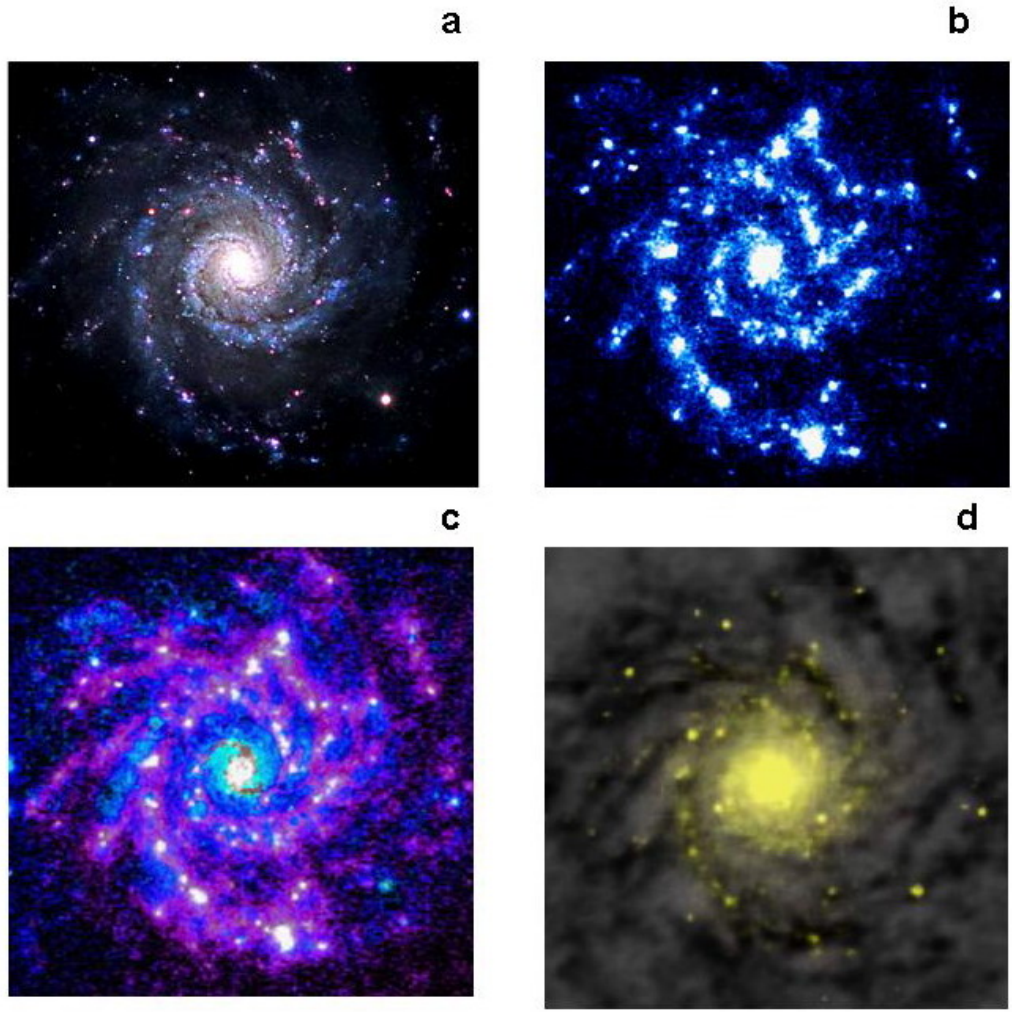

**Figure 11.** M74 (NGC 628) galaxy. The north is at the top, and the east is to the left-hand side.

(a) The optical image*. No dust lane is seen upstream of the eastern (longer) arm, but such a lane is well noted upstream another, shorter arm, which hosting no complexes.

(b) The UIT image**. Note the regular chain of large complexes along the eastern outer arm.

(c) The UIT image superimposed on the *Spitzer* telescope 24-micron image. The lane of the warm dust, connecting the star complexes, is seen now, like it is the case in the NW arm of M31.

(d) The DDS B (yellow) image superimposed on the HI image, prepared from data of Walter et al. 2008).

*) Paul Mortfield and Deitmar Kupke/Flynn Haase/NOAO/AURA/NSF.

**) http://antwrp.gsfc.nasa.gov/apod/ap960409.html.

The semi-regular spacing of the large gas clouds is described in the model of cloud agglomeration in the spiral shock, where resulting stellar aggregates are unbound (Dobbs 2008). This may be the case for the chain of OB-associations in M51 at North of Fig. 12; they are spaced in quasi-regular manner a bit downstream the strong dark lane (Bastian et al. 2005). These star aggregates are clearly too small to be called star complexes.

The genuine complex of ~ 450 pc in size (called G1 in Bastian et al. 2005) is shown in Fig. 12d. The complex contain at least a dozen star clusters; note also the cluster of star clusters near it (to the West). The density of stars and clusters increases to the complex center, and this may indicate it is bound. Note that the strong and wide dust lane (which coincides with the strongest CO lane) disappears just before G1 complex; such a lane is also absent upstream three younger complexes, which are in the SW corner of Fig. 12a,b,c. Narrow and short dust features which are seen near the southern complexes hardly may indicate a spiral shock wave here. It is worth to note a regular spacing of six star complexes and HII regions in the SW segment of the spiral arm, starting with G1 complex (Fig. 12c).

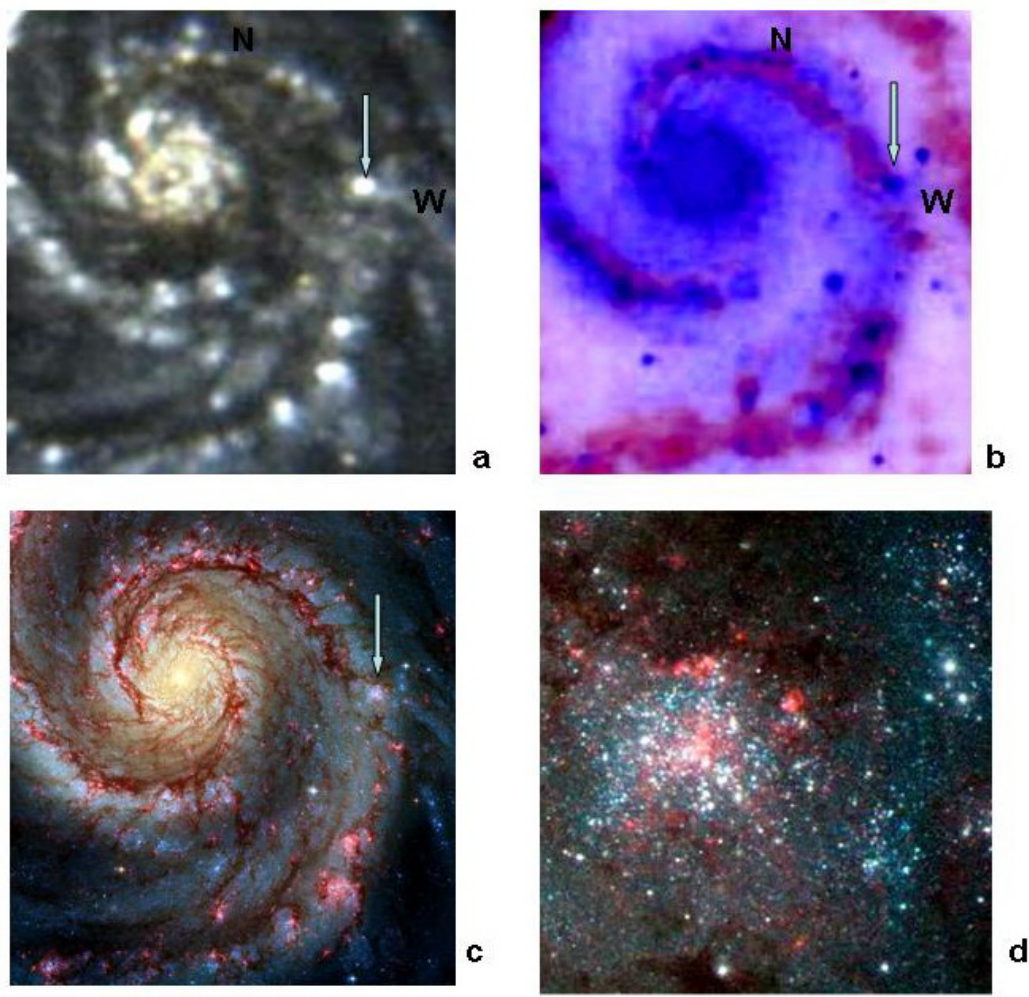

**Figure 12.** Central part of the M 51 galaxy and the star complex G1 (marked by the arrow).
(a) The GALEX space telescope image.
(b) The DSS B image (blue) superimposed on the HI (red) image, prepared from Walter et al. (2008) data
(c) The HST ACS image. The strong dust lane upstream the stellar spiral arm disappears near the complex G1. Note a long spur started from this complex. Unusual abundance of star clusters is seen in the spur in the HST ACS image.
(d) Detail of the Fig. 12c: the G1 complex. It is the best example of the genuine star complex, - with the size about 400 pc, a few small HII regions and notable substructure. It is nor an OB-association, neither a giant HII region ("super-association").

Generally, the HST images of M51 galaxy confirm the impression that star complexes and strong dust lanes along a spiral arm mutually avoid each others. Absence of star complexes in M81, the classical galaxy with the signatures of the spiral shock waves (see, however, Tamburro et al., 2008), points to the same direction. A shock wave in a spiral arm segment seems to be incompatible with possibility of a complex chain formation there .

# 7 DISCUSSION

We have found that in M31 the regular magnetic field is observed in the arm segment which contains the most regular chain of star complexes, and spacing of the latter is close to the half-wavelength of the wavy magnetic field. This demonstrates the tight connection between the regular spacing of complexes and the regular magnetic field along a spiral arm. In this arm segment there is definitely no a spiral shock wave.

It may be a general rule that regular chains of complexes emerge in spiral density waves only if they have a regular magnetic field, the condition for which is a rather low star formation rate, like it is the case for M31. This might explain why the regular chains of complexes in an arm are so rare, much more rare then the spiral shocks. The formation mechanism of such chains is surely the Parker–Jeans instability, as was suggested in the first study where such chains were discovered (Elmegreen & Elmegreen 1983, Elmegreen 1994).

It seems that "beads on a string" morphology and signatures of a spiral shock never meet in the same segment of an arm. The spiral shock can probably prevent the fragmentation of the arm segment into superclouds (parental for star complexes), because it rapidly leads to a high gas density in the entire segment and, as a result, to star formation everywhere in it. In contrast, if there is no fast ubiquitous star formation, the conditions for the slower development of a large-scale instability are met along the gaseous arm and it fragments into superclouds. Initially only within the latter the gas density, sufficient for star formation is reached – in such a way the star complexes may emerge.

Anti-correlation between the shock wave signature and regular spacing of complexes may also be explained with Dobbs & Price (2008) conclusion that spiral shocks generate an irregular magnetic field. Within our above suggestions, this conclusion might be considered one more argument for the regular magnetic field as a precondition to form complexes with regular spacing.

Otherwise, a low rate of star formation (incompatible with a spiral shock) is probably needed to create the regular magnetic field along an arm. Beck (1991) found that the regular magnetic field in NGC 6946 galaxy forms magnetic arms that are located between the gaseous–stellar arms and have the same pitch angle. He explained this as a result of a high star formation rate in the stellar–gaseous arms of NGC 6946, which causes the magnetic field lines to be tangled between numerous chaotically spaced HII regions. Magnetic field is strong over all the galaxy, but is regular only outside the spiral arms, where these regions concentrate.

Chyzy (2008) offered a similar explanation for the anticorrelation between the star formation rate and the degree of magnetic field regularity that he found in NGC 4254. Note that the star formation rate in M31 is relatively low over all the galaxy. Note also the impossibility to divide the spiral arms of NGC 6946 into separate star complexes; they are entirely filled with HII regions and high luminosity stars. The only *bona fide* complex in this galaxy is outside spiral arms (Efremov et al. 2007). It is quite peculiar, having the sharp semi-circular edge and hosting the young supercluster.

The Elmegreen & Elmegreen (1983) list is still the only published result of systematic searches for string of complexes in spiral arms. Note that in 15 of their 22 galaxies the regular strings of complexes are seen in one arm only, the best example being NGC 628. Is it due to the difference in the magnetic field properties? Amongst the galaxies of this list, the most impressive chains of complexes are in the outer parts of the NGC 895 arms, and no dust lanes are seen in these parts. In NGC 5248 the chain of complexes is seen in one arm only, whereas in another arm there are associations and dust lane upstream. Anyway, a dust lane at a galaxy outskirts might be unseen against the low-brightness background (NGC 5033 case?). Amongst galaxies outside the above list, we might note the chains of complexes without a strong dust lane upstream in NGC 5371, NGC 4535 and NGC 4321.

Deeper large scale images of the grand design spiral galaxies are needed, especially in UV and far UV bands, as well in far IR region – to see the warm dust lane far from the center of a galaxy. Anyway far from a center, one hardly may expect existence of a spiral shock wave.

## 8 SUMMARY

Star complexes are the largest units of coherent star formation. Within the flocculent galaxies the complexes larger than the gas disk thickness are sheared and called then the flocculent spiral arms (Elmegreen & Efremov 1996). The existence of a preferential, physically determined average size, distinguishes the genuine star complexes (formed by a large-scale instability) from all of other classes of stellar groups. The sizes of such complexes should be distributed according the normal (Gaussian) law, like it is the case for most complexes in M31 (Efremov 1995, Fig. 9). Note also that D.Elmegreen et al. (2006) have found the Gaussian distribution of sizes of complexes in the (tidal) arm connecting NGC 2207 and IC 2163.

Probably, a lower shear within a spiral density wave let complexes to stay bound while they are inside such an arm. Existence of the preferred mass (and therefore size) for the genuine star complexes may be explained by their formation under action of the large-scale instabilities developing along a GD arm, and especially if this is the magneto-gravitational instability (as it was first stated in Elmegreen, 1982).

The quasi-regular spacing of large gas clouds is also obtained in a model of the agglomeration of smaller clouds in the spiral shock wave, the addition of gravitational instability leading to more regular spacing of more massive clouds (Dobbs 2008). The young star aggregates located at semi-regular spacing within the shocked arm segments are observed indeed (e.g. M51 and NGC 2997 cases), but these aggregates are surely smaller and younger than genuine star complexes. Also, we would like to stress again conclusion of Dobbs & Price (2008) that spiral shocks generate an irregular magnetic field. This may well be a natural explanation for anti-correlation between spiral shock signatures and presence of regularly spaced complexes, provided that such a "beads-on-a-string" morphology of an arm segment is due to the Parker-Jeans instability.

Another indirect evidence for the latter is the very rarity of such a morphology in spiral arms. It is compatible with suggestion that a rather specific mechanism is involved in formation of complexes/superclouds chains. The spiral shocks. are quite common, whereas the regular magnetic field along an arm is not.

## ACKNOWLEDGMENTS

I am very grateful to Bruce Elmegreen for many valuable discussions. Thanks are due to Emilio Alfaro, Elly Berkhuijsen, Arthur Chernin, Dmitry Sokoloff, and Anatoly Zasov for their comments on some issues. I am much indebted to the referee Preben Grosbol for many constructive remarks, which were taken into account in this revised text. The possibility to use the images available at *GALEX, Spitzer, Hubble* and *Swift* space telescopes sites is greatly appreciated. This investigation would be impossible if the ADS were not existing.

The work was supported by the Russian grant for support of leading scientific schools (NSh-433.2008.2).